# SECURING ELECTRONIC TRANSACTIONS TO SUPPORT E-COMMERCE

Mohammad Nabil Almunawar

Faculty of Business, Economics & Policy Studies
Universiti Brunei Darussalam

e-mail: nabil.almunawar@ubd.edu.bn

**Abstract**

*Many reports regarding online fraud in varieties media create skepticism for conducting transactions online, especially through an open network such as the Internet, which offers no security whatsoever. Therefore, encryption technology is vitally important to support secure e-commerce on the Internet. Two well-known encryption representing symmetric and asymmetric cryptosystems as well as their applications are discussed in this paper. Encryption is a key technology to secure electronic transactions. However, there are several challenges such as crytoanalysis or code breaker as well as US export restrictions on encryption. The future threat is the development of quantum computers, which makes the existing encryption technology cripple.*

## Introduction

The growth of the Internet has dramatically changed the way in which data is managed, accessed and used commercially to conduct business electronically. Electronic commerce (e-commerce) is not new; however, recent rapid development of the Internet is surely responsible for the popularity of e-commerce. The new way of commerce through the Internet creates vast opportunities, but at the same time, this new way of commerce poses challenges.

The Internet is a global network, which contains many networks. Therefore, the Internet is a network of networks. Connecting a business to the Internet implies a global reach. In other words, a company can reach anyone who has an access to the Internet such as customers, suppliers, on-line banks, mediators, etc. At the same time, the company can be reached by anyone. As mentioned above, the Internet creates vast opportunities for businesses, but at the same time poses some threats, which if they are not taken care properly will destroy businesses. For example, anybody from anywhere on the Internet (an intruder or a hacker) can illegally enter a company computer resource and messes the computer resource from a remote site. In addition, it is not difficult task to tap a message in the middle of the net and steal or change its content, which is surely a very serious matter.

While the Internet is dramatically changing the way business is conducted, security and privacy issues are of deeper concern than ever before. The Internet is basically an insecure communication medium. Hawker (Hawker, 2000) states that the *only assumption which can safely be made when considering the Internet as a communication medium is that it offers no security whatsoever.*

Most people are skeptical about the security of the Internet. People are happy using the World Wide Web for browsing, finding, reading or downloading information from the Internet. However, when considering sending a credit card number over the Internet, they are reluctant to do it, even if they are told that the transfer is secured. This is because many media expose bad news about the Internet security, although security technology for the Internet exists and good enough for protecting transactions via the Internet.

In addition, a primary fault in evolutionary electronic commerce systems is the failure to adequately address security and privacy issues; therefore, security and privacy policies are either developed as an afterthought to the system or not at all. While the security is actually a major concerned in business transactions over the Internet.

The core activities of e-commerce are business transactions between two parties or possibly mediated by a third party. In fact, the practice conducted by company before the term e-commerce appears is Electronic Data Interchange (EDI), which is basically electronic transaction via computer networks. The major concern of electronic transactions is how to protect transactions from eavesdroppers (which can steal and modify the information in the transactions) and how to make sure those transactions are authenticated. In this paper, we limit the discussion to these issues.

This paper is organized as follows: the next section we discuss the development of the Internet including its security. Some current development of e-commerce on the Internet will also be discussed in this section. In Section 3, we explore encryption technologies to secure electronic transactions. We discuss the current widely used encryption technology on the Internet in Section 4. In Section 5, we highlight challenges on encryption technology and finally section 6 is the conclusion.

## The Internet and E-commerce

The Internet is the world's single biggest networked community; it spans every continent, thousands of networks, millions of computers, and hundreds of millions of people. The Internet began as a research project, the Arpanet, to explore a technology called packet switching, intended to permit robust communication. Before 1990, the growth of the Internet was rather slow. The main users at this time were, research institutions, universities and governments. In addition, most applications of the Internet were text-based and hardly user friendly. This implied that users needed some time to learn those applications to be useful for them.

In 1990, Tim Berners-Lee, a physicist at a laboratory for particle physics in Geneva, Switzerland, developed a new technology to link computers at different places through hyper documents (hypertexts) as spider's web and provided user-friendly navigation. Since the technology can link computers globally through the Internet, it is called the World Wide Web (WWW) or simply the Web.

Shortly after the Web technology released publicly, the growth of the Internet has dramatically increased. For example, in January 1993 there are 1,313,000 host computers connected to the Internet; this number became 12,881,000 in January 1996 and 72,398,092 in January 2000. After the Web was widely adopted, the growth of web server has dramatically increased (from 60,374 in 1995 to 28,125,284 in February 2001!). Table 1 and Figure 1 show the growth of web server based on Netcraft web server survey[1]. Netcraft web survey also shows that the majority of web servers are commercial web servers (dot com), which indicates that the Web is considered as a good medium for commercial purpose, including e-commerce.

Table 1. Web server growth

| Year | Total number of web servers | Number of dot com web servers |
|---|---|---|
| 1995 | 60,374 | 36,529 |
| 1996 | 603,367 | 375,697 |
| 1997 | 1,681,868 | 1,022,224 |
| 1998 | 3,689,227 | 2,095,194 |
| 1999 | 9,560,886 | 5,359,956 |
| 2000 | 25,675,581 | 14,566,664 |
| January, 2001 | 27,585,719 | 15,836,053 |
| February, 2001 | 28,125,284 | 15,855,711 |

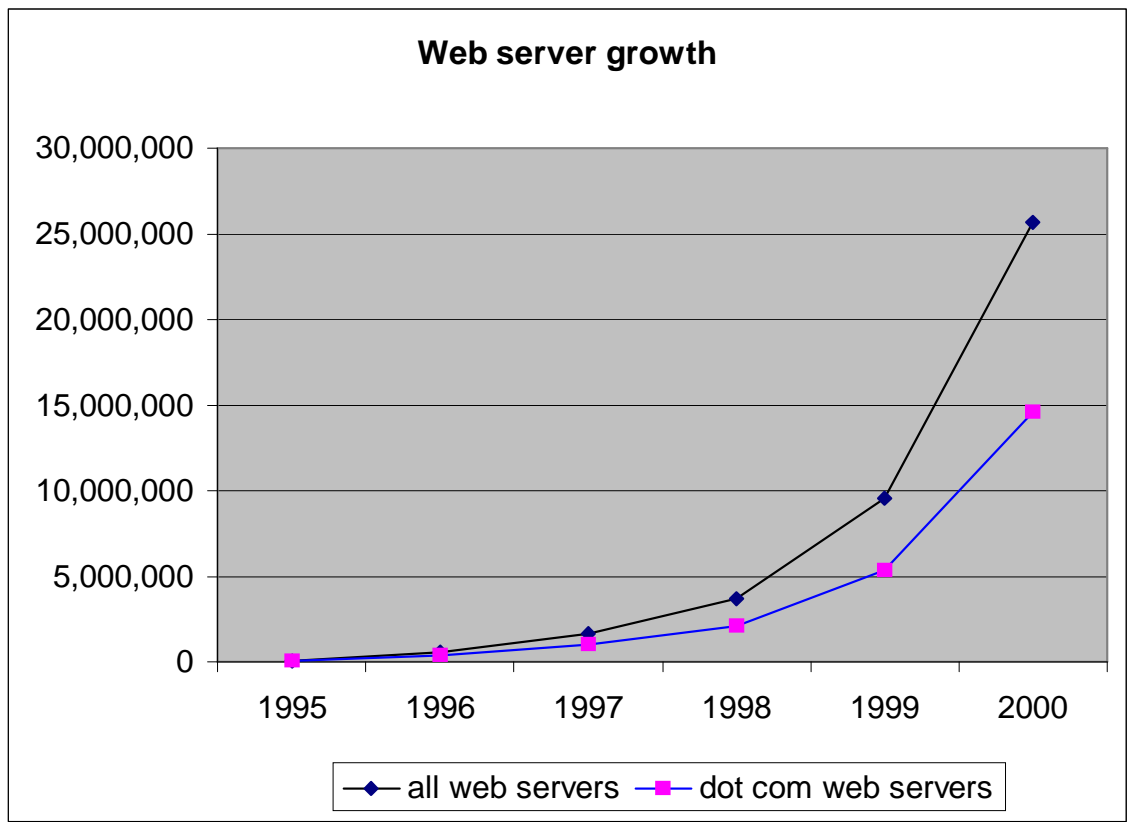

Figure 1. Web server growth

The growth of Internet population is amazing. By June 1996 there were approximately 75 million people (Hoffman and Novak, 1997) worldwide had Internet access. In February 1999 there were approximately 153.5 million people online. This number quickly doubled a year later (304.33 million in March 2000). Yet, within the same year, eight months later (November 2000), the number of people online were 407.1 million, an increase of more than 100 million. The Internet population grows approximately 6.5% per month[2]!

The fast development of the Internet and the Web obviously affects the way people doing commerce electronically or e-commerce. E-commerce exists before the Web, even though the term e-commerce is just widely used recently. Business-to-business electronic transactions have been carried out using private networks or value-added networks. Financial institutions such as banks use private networks to provide electronic transactions to their customers that can be accessed regionally or even globally such as *Plus* and *Cirrus* networks. Prior to the introduction of the Web, most e-commerce activities were carried out using Electronic Data Interchange (EDI). Even now EDI is still being used for conducting transactions, especially business-to-business transactions.

Recently, the Internet (through the Web) is gaining popularity in conducting e-commerce. Table 1 shows that commercial web sites (the sites that end with .com) are the majority of sites on the Web. In addition, the growth of commercial web sites is faster than other any category of web sites, which indicates the future web-based e-commerce will play an important role in the world economy. The web greatly enhances e-commerce, since the Web technology provides user-friendly features, global access, a cheaper way in doing business, and enables more diverse business activities.

Even though the current Internet-based e-commerce is still relatively small compared to the overall e-commerce conducted using private networks, recent market research survey forecasts significant growth within the next few years.

In term of the volume of transactions, business-to-business (B2B) transactions dominate e-commerce activities. For example, according to the US Census Bureau, B2B sales may have accounted for 90 percent of all e-commerce activity in the US in 1999. A release from Nua Internet Survey, March 15, 2000[3], based on research firm Gartner, B2B e-commerce will reach USD8.5 trillion in 2005. This release also shows a significant increase of B2B sales from 1999 to 2000 (in 2000 B2B market was worth USD433 billion, 189 percent increase from 1999). Forecast from Gartners indicates fast growing of B2B sales. It is expected e-commerce sales will be USD919 billion in 2001, followed by USD1.9 trillion in 2002. The market will increase to USD3.6 trillion in 2003, and B2B sales will be worth USD6 trillion by the end of 2004.

Most of the B2B transactions were not conducted through the Internet, even though the Internet provides low cost transactions since companies involved in e-commerce had invested for EDI. In the near future some of these companies will continue to use EDI since switching to the Internet-based solutions mean new investments for them.

Now, let us turn to business-to-consumer (B2C) e-commerce. B2C e-commerce is still small compare to B2B. However, B2C is expected to grow fast in line with the fast growth of the Web. Unlike B2B e-commerce, which used and will continue to use EDI through private networks, B2C will significantly benefited from the growth of an open network such as the Internet. There are several factors that make the Internet (especially the Web) as an attractive medium for conducting e-commerce. For companies, the web is an alternative medium for marketing. The web can be used to disseminate product catalogues, product information, and new products and at the same time advertise those products. In addition, the Web is an effective and efficient medium for a company in providing customer service and support.

Transaction cost over the Internet is lower than over private networks. Also, Connection cost to the Internet is rather cheap. This is the reason of the explosive growth of the Internet population. Of course the global reach of the Internet is one of the most attractive factor for doing e-commerce. Companies can reach customer directly (without intermediaries), hence can reduce distribution cost, which in turn reduce the price of the product/service to customers. At the same time customers have vast opportunities to evaluate products/services by comparison. The comparison can be done quickly through the Web before deciding to purchase them. Technology that allows consumers doing product comparison is now available.  Agent technology provides a capability to automatically visits relevant Web sites, checks information, and selects the best offer. Perhaps convenience is the most attractive feature of web-based e-commerce from customer perspective.

While conducting e-commerce through private networks is relatively secure since the networks cannot be accessed by external entities, conducting e-commerce using an open network such as the Internet carries some risks, especially on transaction's security. Security concerns appear when sellers and buyers negotiate and make contract then the payment take place afterward.

A recent survey in the US (March 2001) from Computer Security Institute (CSI) backed by FBI reports a significant increase in computer crime and information security breaches.  Regarding financial losses, there was an increase of 42.2 percent of losses are reported compare to the previous year and an increase of 214.2 percent compare to the average annual lost over the three years prior to 2000.  Most serious financial losses occurred through theft of proprietary information and financial fraud.[4]

In traditional commerce, negotiations and making contracts are conducted by physical existence of buyers and sellers. Both parties can easily authenticate one another since most likely they know one another.  In e-commerce the issue of authentication is vitally important since both parties do not meet physically.  Someone from somewhere in the Net technically can pretend as sellers or buyers. So, how to make sure that a buyer or a seller is the authentic one? Next, how to represent signatures of contract between buyers and sellers as well as how to protect agreed contracts from modifications by sellers, buyers or perhaps someone else who is capable to do so in the network?

A close related issue is payment.  There are many reports that credit card numbers have been stolen. Therefore, if this issue cannot be properly addressed then the future of internet-based e-commerce is under suspect.  There are better alternatives for payment other than a credit card, however this is beyond the scope of this paper.

Fortunately, the security of e-commerce has been addressed and in fact, there are some commercial security products that can be used to secure transactions through the Internet.  Most of the security products use encryption technology to secure transactions over the Internet such as RSA public key. We will discuss encryption technology to protect transaction security over the Internet in the next section.

**Encryption Technology**

Encryption is a very old technology for keeping messages secret from unauthorized access. One of the oldest methods of encryption was developed by Spartan generals around the fifth century of BC (Hawker, 2000). The basic idea of encryption is only an authorised person can reveal information from an encrypted message by using a key.

To encrypt a message, the message is passed to a method or an algorithm that transforms the message using a *key* (*cryptographic key*) into a meaningless script called *cryptogram* or *ciphertext*. The *ciphertext* can be sent to the intended recipient. The recipient must have a *key* to decrypt the *ciphertext* back to the original message. Figure 2 illustrates the encryption-decryption process.

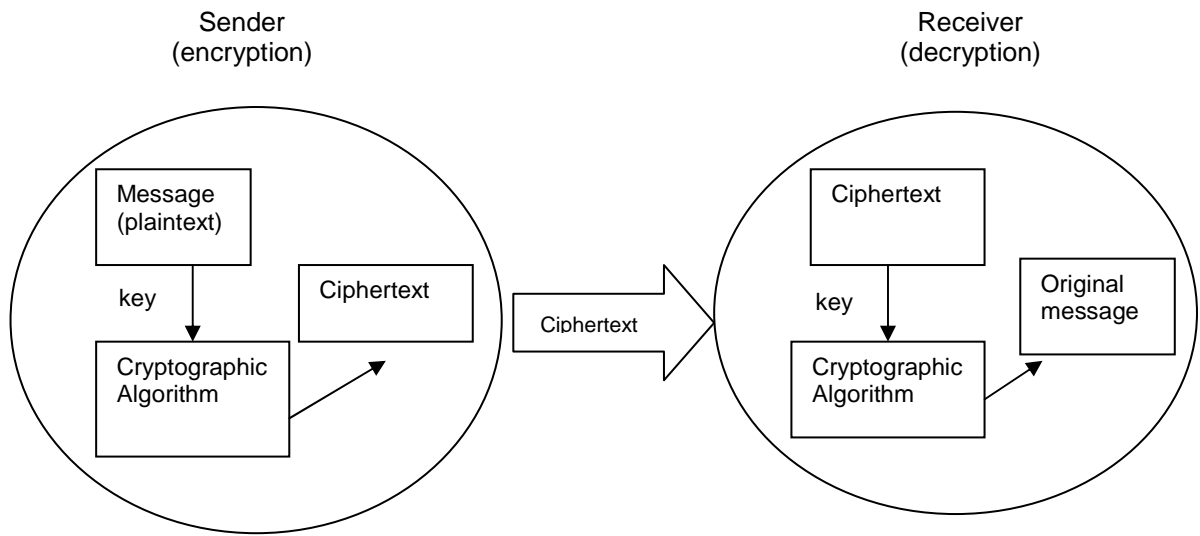

Figure 2: Encryption-Decryption Process

Initially, the use of encryption technology is limited to governments, especially the military. In recent years, the development of encryption algorithms, availability of high computing power to run the algorithms, as well as the need to protect business and commerce transactions has made the encryption technology to spread to business communities. However, highly secure encryption technology products cannot be easily disseminated since governments where the technology developed (such as the US) impose some restrictions on the use of the technology.

Encryption algorithms can be classified into two types, namely *symmetric cryptography* (single key cryptography) and *asymmetric cryptography* or public key (two keys) cryptography. A well-known symmetric cryptography is DES (Data Encryption Standard) developed by IBM for the US government and a well-known public key cryptography is RSA cryptosystem. We will discuss DES and RSA cryptosystem shortly.

**Data Encryption Standard**

Data Encryption Standard (DES) is the most widely used symmetric cryptography. DES was adopted by NIST (National Institute of Standards and Technology) in 1977 to provide an encryption algorithm to be used in protecting federal unclassified information from unauthorised disclosure or undetected modification during transmission or while in storage (NIST, 1995).

The DES algorithm uses a 56-bit key to encrypt *plaintext* to *ciphertext* or to decrypt *ciphertext* to *plaintext*. The DES consists of 16 "rounds" of operations that mix the data and key together in a prescribed manner using the fundamental operations of permutation and substitution. The goal is to completely scramble the data and key so that every bit of the *ciphertext* depends on every bit of the data plus every bit of the key.

The encryption-decryption process can be executed using software or hardware. There are quite number of hardware products such as smart cards and PCMCIA cards that implement the DES algorithm. The main reason to implement the DES algorithm in hardware is the speed of encryption/decryption process, which is much faster than the software implementation since the DES algorithm is complex and computationally intensive. Recently, however, software implementation of the DES algorithm is quite common such as used in protecting passwords or personal identification numbers (PINs). In addition, software implementation is cheaper and easier to install than the hardware one.

According to a NIST report (NIST, 1995). DES has been evaluated by several organisations and has been found to be mathematically sound. The standard must not be modified since according to some analysis the DES algorithm would not be secure if a particular change were made (e.g., if fewer "rounds" were used).

As mentioned above, DES uses a 56-bit key, which means there are 76 quadrillion ($2^{56}$) key possibilities. Of course creating a 56-bit key manually is tedious so the system that implements DES automatically picks a key. Thus, in general the DES encryption is quite secure since attackers have to test billion keys a second in order to find the key in about a year. According to Simonds (Simonds, 1996), the DES algorithm is the most thoroughly tested algorithm ever, and no major weaknesses have been found since its debut in 1977. Bruce and Dempsey (Bruce and Dempsey, 1997) confirm what Simonds stated above.

Simonds (Simonds, 1996) mentions that a study of DES found that a DES system can be broken in ten years. In general, the degree of security of an encryption algorithm, including DES, is dependent on the length of key. One way to strengthen DES security is to repeat the process of encryption several times, which also implies increasing the length of key several times. NIST with US Department of Commerce released a new standard (NIST, 1999) called Triple Data Encryption Algorithm (TDEA). TDEA consists of 3 DES keys. Using TDEA, a message will be encrypted three times with three different keys. So the total length of the keys (called key bundle) is 168 bits long, which surely strengthen the security level of DES.

A newly developed symmetric encryption algorithm is the Advanced Encryption Standard (AES). AES is more powerful than DES. The key lengths are 128, 192, and 256 bits. Further detail of AES can be found in (NIST, 2000).

**The RSA Cryptosystem**

The RSA Cryptosystem is an asymmetric cryptosystem developed by the trio: Ronald Rivest, Adi Shamir and Leonard Adleman (Rivest, Shamir and Adleman, 1978). The RSA cryptosystem is based on the principle that if two large prime numbers are multiplied the resulting number is hard to factor back to its original numbers. In the RSA cryptosystem the two numbers are keys, namely private and public keys. A private key must be kept secret, while a public key can be revealed to anyone. Obviously, the RSA cryptosystem is more complex and harder to manage than DES since it involves two keys. However, an inherent benefit will be revealed shortly.

In the RSA cryptosystem, a sender may encrypt a message using his/her private or public key. Let *A* and *B* be two parties that use the RSA cryptosystem and $K_{PA}$, $K_{TA}$, be the public key and the private key for *A*, $K_{PB}$, $K_{TB}$ be the public key, the private key for *B* respectively. Assume that B knows $K_{PA}$ and A knows $K_{PB}$. There are two possible scenarios:

1. *A* sends a message to *B*. Before sending the message, A encrypts the message using $K_{PB}$. Since *A* uses $K_{PB}$ to encrypt the message then only *B* can decrypt the message using $K_{TB}$. This is called the *encryption path* of the RSA cryptosystem.
2. *A* sends a message to *B*. Before sending the message, A encrypts the message using, $K_{TA}$. Next, B decrypts the message using $K_{PA}$. If *B* can decrypt the message using $K_{PA}$, then the message must come from *A*. This is called the *authentication path*, which can be used as a digital signature (the message is digitally signed by *A*). Note that *A* cannot deny (*non-repudiation principle*) that he/she has signed the message since the message can only be decrypted using *A's* public key ($K_{PA}$).

Both scenarios above can be combined to create an authenticated encrypted message or digitally signed encrypted message. For example *A* wants to send a message securely and make sure that only B can read the message. To tell B that the message is genuinely sent by *A*, *A* digitally signed the message. To do this A encrypts the message using $K_{PB}$, and then re-encrypt the message with $K_{TA}$ prior sending the message to *B*. Upon receiving the message, B decrypts the message twice using $K_{TB}$ and $K_{PA}$. Note that only *B* can decrypt the message and at the same time B authenticates that the message comes from *A*.

The RSA encryption algorithm enables transmission of secret information over an open channel without a common secret key previously shared between them. This algorithm is employed for encryption trapdoor one-way functions that can be executed in "short" time, while the execution of their inverse functions takes practically infinite time. The shortcoming of the RSA algorithm is slower to execute compare to DES.

Like DES, the length of the keys for RSA (both for public and private keys) will determine its strength. A Cray 1S supercomputer in mid-1980s was able to factor 70-digit numbers in less than ten hours. In 1988, a 100-digit number was factored using a network of 50 small computers and in April 1994 a team from Bell Communication Research was able to factor a 129-digit number in about eight months using 1600 computers parallel processed over the Internet (Simonds, 1996).

In order to strengthen the security of RSA the length of the keys must be long enough such that it takes ten to hundreds years to break. Therefore the founder of RSA (Rivest) recommends to use keys between 150 to 230 digits. For example, the length of keys for commercial implementation suggested are 154 digits (or 512 bits) (Simods, 1996).

**Secure Transaction Protocols Used in Current E-commerce**

Electronic transactions are the main activities of e-commerce. Electronic transactions here are nothing more than exchanging information or messages between parties involved in transactions. Messages can be anything such as order form, confirmation messages, credit card numbers or documents. Since transactions are actually the main activities of e-commerce, then it is important to secure the transactions from any kind of threat. As stated by Cameron that the most critical factor in success of electronic commerce is transaction security (Cameron, 1997).

The Internet offers no security whatsoever to business transactions. Information travels over the Internet through series of routing, which means information can be routed through many computer systems before it reaches the trusted server. Any one of these computer systems can represent an opportunity for the information to be accessed or even changed. To support secure information exchange (including e-commerce transactions) currently there are several security protocols based on encryption methods discussed in the previous section. In this section, we will discuss two security protocols commonly used in the Internet to support Internet-based e-commerce, namely Secure Socket Layer (SSL) and Secure Electronic Transaction (SET).

### Secure Socket Layer

Secure Socket Layer (SSL) is a security protocol developed by Netscape Communications to protect communication over the Internet. Although SSL supports many applications such as FTP, Gopher, NNTP (Usenet news), its most popular use is to secure HTTP (the Web)[5]. When SSL is used to secure HTTP, it assures a Web user that he/she communicates with his/her intended Web server and then sends or receives messages securely. To do this, SSL uses the RSA cryptosystem offered by RSA Data Security Inc.[6] for authentication and encryption.

SSL works to protect the Internet communication by the following features[7].

- Server authentication
- Encryption
- Data/Message integrity

To start a TCP/IP (the Internet protocol) connection between a web browser (client) and a secure web server, SSL engages a security agreement (handshake). The handshake enables the client and server to agree on the level of security to be used. The handshake includes checking the server's digital certificate. Digital certificates are electronic files (containing user name, user's public key, and name of Certification Authority (CA) issuing the certificate and other attributes) that act as an online passport. Digital certificates are issued by trusted third parties known as certificate authorities such as VeriSign. Therefore, a digital certificate does connection verifications between the server's public key and the server's identification. If this initialization process is successfully done all data transmission between the client and the server are encrypted by the RSA cryptosystem.

SSL support two types of key, 40-bit and 128-bit long (called session keys). As we discussed in the previous section, the longer the key the stronger encryption will be. For instance, 128-bit sessions are trillions of times stronger than 40-bit sessions Most browsers support 40-bit SSL sessions, and the latest browsers, including Netscape Communicator 4.0, enable users to encrypt transactions in 128-bit sessions. Unfortunately due to export restrictions any browsers outside the US and Canada must use 40 bit session keys, whereas the US and Canada use 128 bit session keys.

In summary, the SSL security protocol provides data encryption, server authentication, message integrity, and optional client authentication for a TCP/IP connection. Because SSL is built into all major browsers such as Netscape navigator or Microsoft Internet Explorer and web servers, simply installing a digital certificate turns on their SSL capabilities[8].

### Secure Electronic Transaction

Secure Electronic Transaction (SET) was jointly developed by Visa and MasterCard and supported by GTE, IBM, Microsoft, Netscape, RSA, SAIC, Terisa, and VeriSign. The SET's main objective is to provide secure payment using credit cards over an open network such as the Internet.

SET protects payment information based on authentication (merchants & cardholders authentication) and encryption of payment information, which is basically similar to SSL. How SET protects payment using credit cards is described in (SETCo, 2001) as follows:

- First, it enables a cardholder to authenticate that a merchant is authorised to accept payment cards in a secure manner using SET technology.
- Second, it enables a merchant that is using SET technology to authenticate the payment card being used in the transaction.
- Third, SET technology uses an advanced encryption system to protect personal payment information during transfer over the network.
- Fourth, SET technology makes sure only the intended recipient reads the payment information. Information which can only be decoded by a merchant and a financial institution that both use valid SET technology.

According to the SET specification, SET uses both symmetric and asymmetric cryptosystems to protect transactions on the open networks (SETCo, 1997). A typical secure transaction using SET scheme in B2C e-commerce such as on-line shopping can be described as follows:

Both parties (a cardholder and a merchant) exchange their public keys through the digital certificate issued by a Certificate Authority during initialisation process.

Cardholder site:

1. The cardholder encrypts the payment information using his/her private key, which also means that he/she has digitally signed the payment.
2. The cardholder re-encrypts the digitally signed payment using a randomly generated symmetric encryption key to ensure message confidentiality.
3. Finally a cardholder encrypts the message (from step 2) with the merchant's public key, creating a secure "digital envelope" and send it to the merchant.

Merchant site:

1. The merchant opens the "digital envelope" using the merchant's private key. Note that only the intended merchant can open the envelope
2. The merchant checks the cardholder's digital signature by using the cardholder's public key.
3. The merchant decrypts the payment information using the symmetric key attached by the cardholder.

The above scenario allows a merchant to access a cardholder's credit card number. To hide a credit card number from a merchant the SET protocol uses a payment gateway (belong to a credit card processor). In this scheme a cardholder's order information will be encrypted using a merchant's public key and his/her credit card detail will be encrypted using the payment gateway's public key and sent to a merchant. Hence the merchant can only open the order information and will pass credit card detail to the payment gateway for verification.

**Security Challenges**

The major challenge of encryption-based security is cryptoanalysis, an activity to break the encryption by guessing keys. Using the current methodology of encryption, a *ciphertext* can be decrypted by a cryptoanalist. The question is how long can an encryption be broken. The strength of an encryption can be measured by how hard the encryption can be broken, which depends on the encryption algorithm, the length of the key and the power of the computer(s) used to crack the encryption.

Attacks on security based on cryptosystems have been published in many media. DES has been designed as a relatively secure encryption method, however powerful computers or networks of computers can break it in minutes. Table 2 shows some estimates to break DES.

Table 2. Estimates of Breaking 56-bit DES[9]

| Type of Attack | Budget (US $) | Time to crack the key |
|---|---|---|
| Pedestrian hacker | 400 | 38 years |
| Small business | 10,000 | 556 days |
| Corporate department | 300,000 | 3 hours |
| Large company | 10,000,000 | 6 minutes |
| Intelligence Agency | 300,000,000 | 12 seconds |

RSA Data Security has offered challenges to public to break DES since 1997. In 1999, a team using a network of 100,000 PCs on the Internet was able to break DES key in 22 hours and 15 minutes. See Table 3 for detail.

Table 3 summarizes the RSA Code-Breaking contest (breaking a 56-bit DES key)[10]

| Year | Winner | Time to break |
|---|---|---|
| January 1997 | a team led by Rocke Verser of Loveland, Colorado | 96 days |
| February 1998 | Distributed.Net | 41 days |
| July 1998 | Electronic Frontier Foundation (EFF) | 56 hours |
| January 1999 | A worldwide coalition of computer enthusiasts, worked with the Electronic Frontier Foundation's (EFF) "Deep Crack," a specially designed supercomputer, and a worldwide network of nearly 100,000 PCs on the Internet. | 22 hours and 15 minutes |

Given the current computer technology, the public key cryptosystem is theoretically a very secure system. Factoring a 125-digit number theoretically takes 40 quadrillion years by a powerful computer in 1976. However, in 1994, a 129-digit number was factored by a networked of 1,600 computers in about one month (Cameron, 1997).

A weaker public key (40-bit key) implemented in Netscape Navigator was broken by a French student using a brute force search (try every possible combination) in 8 days using 120 workstations and several large computers in September 1995 (Hawker, 2000).

Note that attempts to break cryptosystems are dedicated to break one key. If a document is encrypted several times with different keys then the encrypted document is harder to break. This is the reason to encrypt a document several times such as the use of three keys in TDEA. Similarly combining two or more encryption methods (such using DES and Public key in SET) will strengthen the encryption.

Another effort to strengthen the security is to use different keys at different times. Although there is an attempt to break a key, if it is success then most likely the revealed key is no longer used, hence void the effort.

The future real challenge of current encryption technology is future computers called quantum computers. Today computers rely on a binary representation (only two states: 0 and 1, called bit) to store and compute data or information. Quantum computers use qubits (quantum bits) to store and compute. Unlike classical bits, qubits can exist simultaneously as 0 and 1, with the probability for each state given by a numerical coefficient. Describing a two-qubit quantum computer thus requires four coefficients. In general, $n$ qubits demand $2^n$ numbers, which rapidly becomes a sizable set for larger values of n. Obviously, a quantum computer is immensely powerful because it uses multiple states at once (called superposition) and at the same time can act on all its possible states simultaneously. Thus, a quantum computer naturally performs myriad operations in parallel, using only a single processing unit (Gershenfeld and Chuang, 1998).

The most famous example of the extra power of a quantum computer is Peter Shor's[11] algorithm for factoring large numbers. Public key depends on factoring being a hard problem. Despite much research, no efficient classical factoring algorithm is known. However, Shor found that a quantum computer could, in principle, accomplish this task much faster than the best classical computer ever could. In other words there exists an efficient quantum algorithm to factor large numbers. Obviously, this discovery had an enormous impact. Suddenly, the security of encryption systems that depend on the difficulty of factoring large numbers become suspect. And because so many financial transactions are currently guarded with such encryption schemes, Shor's result sent tremors through a cornerstone of the world's electronic economy (Gershenfeld and Chuang, 1998).

Quantum computer is still in its infancy, research and experiments on quantum computers is going on, we do not know yet when quantum computers will be available on our desks. However, quantum computers are definitely a strong warning for current encryption technology. A completely new approach on encryption technology will be needed if quantum computers available. Quantum cryptography?[12]

## Conclusion

Transaction security is vital in e-commerce. Hesitation or scepticism in transaction security over the Internet is a crucial issue needs to be taken care seriously. People are happy with the development of the web where they can browse the Internet and find information they need easily. However, when it comes to decide to buy a product/service over the Internet many people worry about the transaction security. Similarly, firms worry about online frauds. According to the study conducted by the World-wide E-Commerce Fraud Prevention Network[13], half of US businesses believe that online fraud is a significant problem, while almost 10 percent say it is "the most significant problem" they face. The study also found that only 1 percent of companies say they "never worry" about online fraud.

Encryption technology discussed in this paper is the key technology to make online transaction over the Internet secure. Of course no one can guarantee 100% security. Fraud exists in current commerce systems: cash can be counterfeited, checks altered, credit card numbers stolen. Yet these systems are still successful because the benefits and conveniences outweigh the losses. Similarly fraud will still exist in e-commerce even though encryption technology is good enough to protect electronic transactions, but at least a good encryption technology can reduce fraud significantly.

## Endnotes

1. Netcraft survey: http:// www.netcraft.co.uk
2. http:// www.nua.ie/surveys/how_many_online/world.html
3. http://www.nua.ie/surveys
4. The CSI report reveals that thirty-five percent (186 respondents) were willing and/or able to quantify their financial losses. These 186 respondents reported $377,828,700 in financial losses. (In contrast, the losses from 249 respondents in 2000 totalled only $265,589,940. The average annual total over the three years prior to 2000 was $120,240,180.)  As in previous years, the most serious financial losses occurred through theft of proprietary information (34 respondents reported $151,230,100) and financial fraud (21 respondents reported $92,935,500), source: http://www.gocsi.com
5. http://www.rit.edu/~esp3641/ssl.html
6. http://www.rsa.com
7. http://www.cs.umbc.edu/~woodcock/cmsc482/proj1/ssl.html
8. www.netscape.com/security/techbriefs/ssl.html
9. Source: Blaze et al., Minimal Key Lengths for Asymmetric Ciphers to Provide Adequate Security in (Cameron, 1997 p. 60)
10. http://www.rsa.com/news/pr/990119-1.html
11. Peter Shor homepage: www.research.att.com/~shor
12. See Daniel Gottesman for detail (http://qso.lanl.gov/~gottesma/Crypto.htm)
13. http://www.merchantfraudsquad.com